\documentclass{sig-alternate-05-2015}

\usepackage{authblk}

\usepackage{caption}
\usepackage{subcaption}
\usepackage{graphicx}

\usepackage{listings}
\makeatletter
\def\@copyrightspace{\relax}
\makeatother

\begin{document}






%

\title{The Effect of Social Feedback in a Reddit Weight Loss Community \titlenote{This is a preprint of an article appearing at ACM Digital Health 2016}}
%
%
%
%
%

\author[1 2]{Tiago O. Cunha}
\author[1]{Ingmar Weber}
\author[3]{Hamed Haddadi}
\author[2]{Gisele L. Pappa}
\affil[1]{Qatar Computing Research Institute, HBKU, Qatar}
\affil[2]{Federal University of Minas Gerais, Brazil}
\affil[3]{Queen Mary University of London, UK}
\affil[ ]{tocunha@dcc.ufmg.br, iweber@qf.org.qa, hamed.haddadi@qmul.ac.uk, glpappa@dcc.ufmg.br}

\renewcommand\Authands{ and }

\maketitle
\begin{abstract}
It is generally accepted as common wisdom that receiving social feedback is helpful to (i) keep an individual engaged with a community and to (ii) facilitate an individual's positive behavior change. However, quantitative data on the effect of social feedback on continued engagement in an online health community is scarce. 
In this work we apply Mahalanobis Distance Matching (MDM) to demonstrate the importance of receiving feedback in the ``loseit'' weight loss community on Reddit. Concretely we show that (i) even when correcting for differences in word choice, users receiving more positive feedback on their initial post are more likely to return in the future, and that (ii) there are diminishing returns and social feedback on later posts is less important than for the first post. We also give a description of the type of initial posts that are more likely to attract this valuable social feedback.
Though we cannot yet argue about ultimate weight loss success or failure, we believe that understanding the social dynamics underlying online health communities is an important step to devise more effective interventions. 
\end{abstract}

%
%



\section{Introduction}

According to the World Health Organization, the prevalence of obesity has nearly doubled in the last 30 years\footnote{\url{http://www.who.int/gho/ncd/risk\_factors/obesity\_text/en/}}, making the condition a major public health problem. Obesity is associated with significantly increased risk of more than 20 chronic diseases and health conditions \cite{thiese:2015}, and directly affects quality of life. 

It is well known that people dealing with obesity can greatly benefit from support groups to helping them lose weight \cite{wang:2014}. In the digital era, an increasing number of people suffering from this condition are turning to online sources, both to collect information and to connect with like-minded people to create virtual support groups. The digital data generated from these online interactions creates new opportunities to study various topics related to obesity, such as the effects of online social feedback \cite{Ballantine:2011}, social capital \cite{Leroux:2013} and social identity \cite{Chiu:2015}.

In this work, we explore data from a Reddit weight loss community and ask the following research questions: \textbf{Q1} - Does social feedback received on the first post have an effect on the probability of users to remain engaged with the weight loss community?, \textbf{Q2}: Is social feedback on later posts as important as in the early posts for maintaining user engagement?, and \textbf{Q3} - What are the characteristics of initial posts that attract more social feedback than average posts?





Reddit\footnote{\url{https://www.reddit.com}} is a social news website and forum. Its content is organized in sub-communities by areas of interest called subreddits. In 2015 it had 8.7 million users from 186 countries writing 73.2 million posts and 725.9 million comments in 88,700 active subreddits\footnote{``Active'' is determined by having 5 or more posts and comments during at least one week in 2015.}. 
For our study we look at the popular weight loss subreddit \emph{loseit}\footnote{\url{http://www.reddit/r/loseit}}. 

In \textit{loseit}, the user-generated content comprises various topics related to obesity and weight loss, such as personal experiences, recommendations and feedback about certain medications, medical procedures, diets or exercises. Last but not least there is also emotional support in the form of encouragement, sympathy, and success stories. For a given post, other users can provide feedback by (i) replying with a comment or (ii) upvoting the original post as a sign of ``liking'' it.
We study whether these forms of social feedback have an effect on a user's propensity to continue to engage beyond their first post. In particular we look at whether for pairs of similar initial posts by new-comer users, users who receive more feedback from the community are more likely to return to the subreddit in the future. 

We found robust evidence that any type of feedback does indeed increase a user's probability to return. Furthermore, there is an effect of ``diminishing returns'', where social feedback appears to be most important for a user's initial post but the support received looses its importance for later posts.

The rest of this paper is organized as follows. In the next section we review related work, in particular works investigating social feedback, user engagement and Reddit. Section~\ref{sec:data} then describes the data collection. Our results are presented in Section~\ref{sec:results}, broken down in three parts. First, Section~\ref{sec:results:pairing} shows the analysis at the heart of our work: applying Mahalanobis Distance Matching and observing if the user who receives more social feedback returns more often than their less fortunate counterpart. 
Second, Section~\ref{sec:results:diminishing} then extends this analysis by looking at whether social feedback is only of importance for the initial post or also for latter posts. 
Finally, in Section~\ref{sec:results:qualitative} we describe general characteristics of posts that are successful in attracting social feedback.


\section{Related work}

The relationship between social interactions, social feedback and health outcomes has been extensively studied in the medical literature. Research has shown that conditions such as smoking, depression and coronary disease, among others, may be controlled if individuals receive enough social feedback in the form of social support \cite{umberson:2010}. In the context of obesity, the influence of social feedback on engagement in healthy eating and physical activity, as well as on achieving successful outcomes in weight reduction programs, has also been demonstrated \cite{wang:2014}. Other studies also showed that there is an independent role of social feedback on health-related quality of life among obese individuals \cite{herzer:2011}.

Encouraging continued user engagement is essential for the success of weight loss programs, previous studies showed that users who stay longer in these programs have greater success in achieving their goals \cite{patrick:2011,Glasgow:2007}.  

The importance of user engagement also holds for online groups, where social feedback has been shown to play an important role in new-comer user engagement \cite{Burke:2009}. However, most observational studies do not control for covariates that can affect the probability of receiving feedback, which limits its ability to explain the effect of the feedback received. As examples of covariates, Althoff \textit{et al.}\ \cite{Althoff:2014} found that posts with linguistic indications of gratitude, evidentiality, and reciprocity are more likely to receive feedback and obtain success while asking for free pizza in the subreddit ``Random acts of Pizza''. In our work we reduce the bias created by such effects by applying Mahalanobis Distance Matching. 



Reddit has been used to study different health conditions under different perspectives, including social feedback. For example, Choudhury \textit{et al.} \cite{Choudhury:2014} analyzed the discourse of Reddit posts and comments looking for indications of depression. They found that Reddit users in certain communities explicitly share information about mental health issues, potentially to gather social feedback. Eschler \textit{et al.}\ \cite{eschler:2015}, in turn, investigated the behavior of patients in different cancer stages in the subreddit r/cancer. From a content analysis of the posts, they concluded that patient and survivor participants show different types of information and emotional needs according to their illness phase, and suggested certain community reorganization to make information access easier for people with cancer in different stages. Tamersoy \textit{et al.}\ \cite{tamersoy:2015} performed an analysis of addiction in Reddit, focusing on tobacco or alcohol. They collected data from two smoking and drinking abstinence communities and identified the key linguistic and interaction characteristics of short-term and long-term abstainers. Then, they built a supervised learning framework based on the characteristics above to distinguish long-term abstinence from short-term  abstinence. 



\section{Dataset}\label{sec:data}
The data used in our analysis covers 5 years (August 2010 to October 2014) and was crawled from Reddit using PRAW (Python Reddit API Wrapper), a Python package that allows simple access to Reddit's official API. In Reddit users can submit content, such as textual posts or direct links to other sites, both collectively referred to as \emph{posts}. The community can then vote posted submissions up (\emph{upvotes}) or down (downvotes) to organize the posts and determine their position on the site's pages. Information on downvotes is, however, not exposed via the API. Users can also reply to posts with \emph{comments}.

\begin{figure}[ht]
	\centerline{\includegraphics[scale = 0.40]{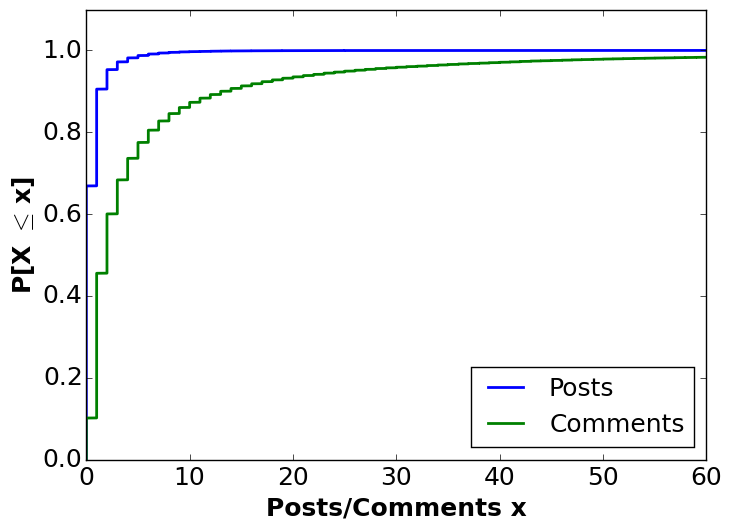}}
	\caption{Cumulative distribution function of number of posts and comments per user.}
	\label{fig:CDFPostsComments}
\vspace{-0.4cm}
\end{figure}

The data we collected include posts, comments and other metadata (i.e., timestamp, user name, number of upvotes). In total, we obtained 70,949 posts and 922,245 comments. These data were generated by 107,886 unique users, of which 38,981 (36.1\%) wrote at least one post and 101,003 (93.6\%) at least one comment. Table~\ref{tab:basicStatistics} shows the mean, median and standard deviation (SD) for basic statistics of the dataset, including the length of posts and comments and the number of daily messages. Figure~\ref{fig:CDFPostsComments} shows the cumulative distribution function (CDF) over posts and comments per user. 

\begin{table}[ht]
\centering
\caption{Basic statistics of \textit{loseit} dataset.}
\begin{small}
\begin{tabular}{|l|l|l|l|}
\hline
							  & Mean & Median & SD \\ \hline
Posts per day &45.5 &45 & 22.7\\ \hline
Comments per day & 586.6 &599 & 264.3\\ \hline
Upvotes per post & 35.7 &6 & 126.7\\ \hline
Upvotes per comments &3.1 &2 & 11.4\\ \hline
Words per posts & 89.3 & 64 & 95.8\\ \hline
Words per comments & 25.5 & 14 & 35.3\\ \hline
\end{tabular}
\end{small}
\label{tab:basicStatistics}
\vspace{-0.4cm}
\end{table}

\section{Results}\label{sec:results}
\subsection{Mahalanobis Distance Matching}\label{sec:results:pairing}
In order to investigate if users who receive more social feedback on their initial post are more likely to engage in the community and return for a second activity the most obvious approach is to split the data into two cohorts: users that receive social feedback on their first post and those that do not and then, for each cohort, compare the fraction of users returning to the subreddit.

Though intuitive, this kind of analysis is strongly limited by the effect of other covariates. For example, a generally optimistic user might write a first post with a more positive tone than a more pessimistic counterpart. Let us then imagine that, in response, the former user receives lots of comments and the latter receives none. Now let us further imagine that the former user returns for more activity on the subreddit later, whereas the latter user is never to be seen again. The question then arises whether the comments received ``caused'' the former user to return or, rather, whether that user was at a higher disposition to return anyway and the social feedback received was a mere correlate.

To obtain a less biased estimate of whether the ``treatment'' of receiving social feedback has an effect, we apply a matching method. Matching is a preprocessing for reducing model dependence. The most commonly used matching method is Propensity Score Matching (PSM) \cite{rosenbaum-1983}, which aims to approximate a complete random experiment. PSM first builds a model to predict the probability of a particular user to receive the treatment. Users are then matched according to their probability of receiving the treatment. However, recently King and Nielsen \cite{king:2015} showed that this method is suboptimal and that PSM can under certain circumstances even increase the bias in the data.

Here we apply the Mahalanobis Distance Matching (MDM) \cite{rubin:2006} which approximates a fully blocked experiment. In MDM first we measure the distance of the users based on the covariates -- in our case the content of the first posts -- then we match each ``treated'' user with the nearest ``non-treated'' user. Next, the analysis looks at whether, once correcting for differences in prior probabilities, the act of receiving the treatment affects the observed outcome.

To define the treated and non-treated groups, we ranked all the first posts by the amount of feedback received, then we considered the top 40\% as the treated group and the bottom 40\% as the non-treated group. The middle 20\% were dropped from the analysis. When matching the users we started with the ``most treated'' ones, i.e., those user in the top 40\% that received the largest amount of feedback, going until the ``least treated'' one. Each user in the treated group was matched with the nearest user in the non-treated group without replacement. To make sure that our method is matching similar users, we only considered pairs with a similarity value greater or equal to 0.8.

The definition of the covariates was motivated by the hypothesis that posts with similar content have a similar probability of receiving feedback. We could include other features to the covariates, such as user attributes. However, due to the anonymous nature of Reddit, user attributes like demographics or profile images are not available in \textit{loseit}, and hence the sole focus on the post's content is natural.

When matching treated and non-treated users, we used a topical representation of their first post's content extracted by Latent Dirichlet Allocation (LDA) \cite{Blei:2003}. The required parameters -- number of topics, number of iterations, $\alpha$ and $\beta$ -- were empirically defined as 100 and 1,000, 1 and 0.1. Then to compute the distance among users, we used the cosine similarity. Table~\ref{tab:similarPosts} shows parts of the most similar pair (treated and non-treated considering the number of comments received) of first posts according to the LDA model. This pair had a cosine similarity of 0.99.

\begin{table}[ht]
\centering
\caption{Parts of the most similar pair of posts.}
\resizebox{\columnwidth}{!}{%
\begin{tabular}{|p{7.9cm}|}
\hline
\textbf{Treated:} I do eat a pretty broad variety of stuff, but I'll start out. I'm a calorie counter. I aim between 1200-1600/day: one poached egg(70 cal), toast(100 cal), 3 strips center cut bacon (70 cal)...\\ \hline
\textbf{Non-treated:} My go-to 100 calorie snacks: 18 grams of peanuts, 20 grams of pepitos (shelled pumpkin seeds)- 10 Marys(TM) Crispy Crackers, 140 grams of 2\% fat greek yogurt...\\ \hline
\end{tabular}}
\label{tab:similarPosts}
\vspace{-0.4cm}
\end{table}



We applied our methodology to the first post of 37,278 users in \textit{loseit}. 
Table~\ref{tab:similarityMatchingApproach} presents the probabilities of a user returning to the community for a second activity according to the amount of feedback received in the first post. We investigated the effect of four types of feedback: all comments (13,385 pairs), only-positive comments (13,441), only-negative comments (13,620 pairs) and upvotes (12,604 pairs). The sentiment of the comments was computed using the Valence  Aware  Dictionary  for sEntiment Reasoning (VADER) \cite{Hutto:2014}. Finally, for the definition of a ``return activity'', i.e., the second activity of a user in the community, we considered both of the following: (i) when the user comes back to create a new post, and (ii) when he comes back to create a new post or comment on an existing post.\footnote{Note that we cannot obtain information for when or if a user \emph{reads} posts on Reddit. Correspondingly we cannot measure the effect of social feedback on the ``lurking'' probability.}
  

As expected, when considering both posts and comments as a user's second activity, the return probabilities increased significantly, e.g., the number of comments received increases from 32.5\% to 76.3\% for the treated users. This is explained by the fact that \textit{loseit} users comment much more than create their own posts (see Figure~\ref{fig:CDFPostsComments}). The fifth row shows the relative difference in probability of receiving more feedback. We found that all types of social feedback have a positive effect to users coming back for a second activity. We ran a chi-square test over the probabilities of bigger and smaller and for all types of feedback, and all the differences were statistically significant. We also found the effect of positive-only comments is bigger than all-comments and that negative-only comments have a smaller but still positive effect, possibly due to noise in the sentiment classifier. For example, the word ``lost'' as in the sentence ``I've lost a lot of weight.'' in the context of \textit{loseit} is very positive, but VADER classifies it as negative.




\begin{table*}[!htbp]
\centering
\caption{Probabilities of a user returning to the community to perform at least one more activity (posts only or posts or comments) based on the amount of feedback (comments, positive comments, negative comments or upvotes) received in the first post.} 
\label{tab:similarityMatchingApproach}
\resizebox{\linewidth}{!}{
\begin{tabular}{|l|l|l|l|l|l|l|l|l|}
\hline                        
							&	\multicolumn{2}{c|}{All-comments} & \multicolumn{2}{c|}{Only-negative comments} & \multicolumn{2}{c|}{Only-positive comments} & \multicolumn{2}{c|}{Upvotes} \\ \hline                        
							& Post & Post or comment & Post & Post or comment & Post & Post or comment & Post & Post or comment\\ \hline
Probability of treated return  & 32.6\% & 76.3\% & 30.6\% & 72.9\% & 33.1\% & 77.4\% & 31.4\% & 74.8\%       \\ \hline
Probability of non-treated return & 26.9\% & 61.2\% & 29.1\% & 67.2\%  & 26.7\% & 60.8\% &  26.9\% & 65.9\%     \\ \hline
P(returns|``treated'')/P(returns|``non-treated'') & \textbf{21.2}\% & \textbf{24.7}\% &  \textbf{5.1\%} & \textbf{8.5\%} & \textbf{24.0\%} & \textbf{27.3\%}  & \textbf{16.7\%} & \textbf{13.5\%}   \\ \hline
Probability of both return    & 9.2\%  & 47.2\% & 9.4\%  & 49.2\% & 8.9\% & 47.4\% & 8.8\% & 49.7\%  \\ \hline
Probability of neither return & 49.7\% & 9.7\%  & 49.6\%  & 9.0\%  & 49.0\% & 9.2\% & 50.4\% & 9.0\%  \\ \hline
\end{tabular}}
\vspace{-0.4cm}
\end{table*}

\subsection{Diminishing returns}\label{sec:results:diminishing}
In the previous section we have presented evidence that social feedback on a user's initial post increases their probability of returning between 14\% and 24\%. However, two questions arise from these results: if there is something special about a user's initial post and how they are ``welcomed'' in the community, and  whether users are in ever lasting ``need'' of social feedback, even once they have become veterans in the community.

To answer these questions, we look for effects of ``diminishing returns'', i.e., if receiving social feedback on a user's later posts gives less of a boost than receiving it earlier. We also look at the amount of social feedback received, as we hypothesize that any social feedback is better than none, but that receiving four comments is only marginally better than receiving three.

\begin{figure}[!ht]
        \centering
        \begin{subfigure}[b]{.47\textwidth}
          \centering
          \includegraphics[width=\textwidth]{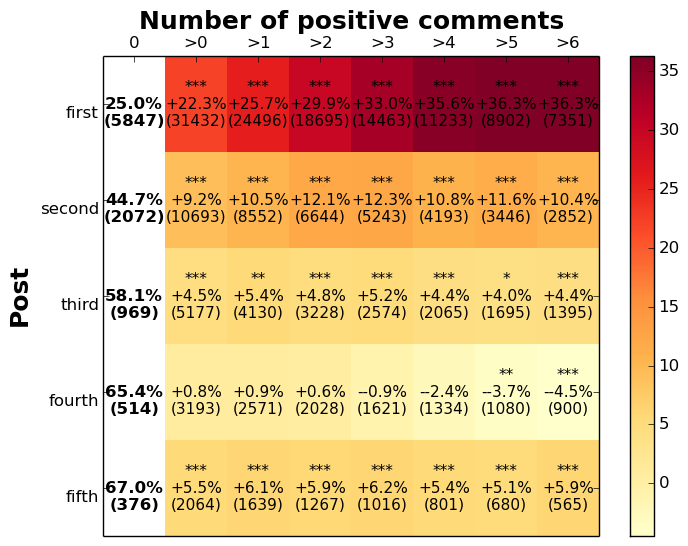}
          \caption{Positive comments.}
          \label{fig:DRNumPosCommentsPosts}
        \end{subfigure} 
         \hfill         
        \centering
        \begin{subfigure}[b]{.47\textwidth}
          \centering
          \includegraphics[width=\textwidth]{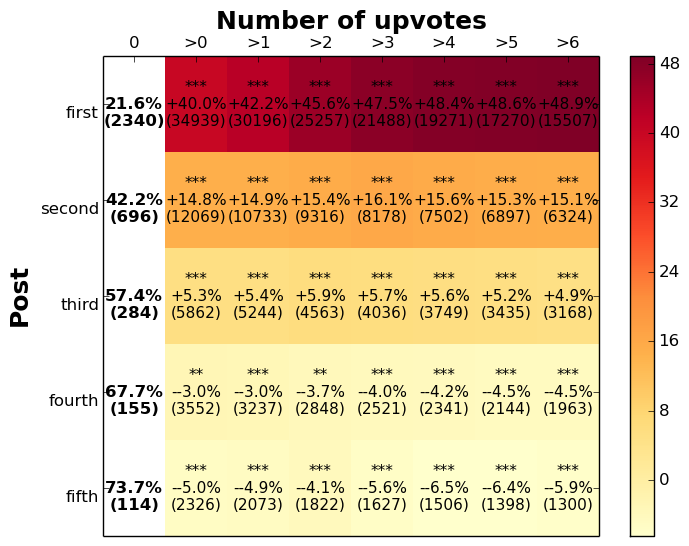}
          \caption{Upvotes.}
          \label{fig:DRNumUpvotesPosts}
        \end{subfigure}  
        \caption{Diminishing returns for the number of positive comments and upvotes.} 
          \label{fig:DRNumPosComments}
\vspace{-0.2cm}
\end{figure}

In Figures~\ref{fig:DRNumPosCommentsPosts} and \ref{fig:DRNumUpvotesPosts} we present the diminishing returns for the amount of feedback received from the first until the fifth post. We show results for both receiving positive comments as feedback, and for receiving upvotes as feedback. The first column shows the probability of a user's later return for activities in the community when receiving no feedback at all. The following columns present the relative difference (compared to receiving zero feedback) of receiving more than 0, more than 1, up to more than 6 feedbacks. For each of these columns, the return probability is also compared for its significance level against the baseline probability of receiving no feedback. The stars indicate the significance levels for a chi-square test for equality, with the number of asterisks corresponding to the p-values, *** for 0.1\%, ** for 1\% and * for 5\%. For all the cells, the number of users considered appears between parentheses.

As is immediately evident from the color coding, receiving social feedback for the first post corresponds to by far the largest relative boost in return probability.  Furthermore, after receiving more than four bits of social feedback the further gains are largely irrelevant. For some rows in the figure we also observe \emph{negative} values indicating a \emph{decrease} in return probability. We hypothesize that this is due to random noise induced by the smaller and smaller user sets for later and later posts.


\subsection{Qualitative analysis}\label{sec:results:qualitative}
So far we have shown that (i) receiving social feedback seems to boost the return probability of a user, and that (ii) this boost shows ``diminishing returns'' and receiving social feedback on later posts is less impactful. One obvious question we have not yet answered though is: what types of post receive social feedback? Put differently, how should newcomers behave to boost their chances of receiving feedback?

We attempt to answer this question through qualitative analysis. We sort all first posts by their number of received comments. Then we compare the content for the top 10\% posts (3,727 posts) to the content of the bottom 10\% posts (3,727 posts). For better clarity we removed the most common words present in both groups (e.g. weight, year, day, time, now, week, calorie and started).

\begin{figure}[!ht]
\centering
\begin{subfigure}{.5\linewidth}
  \centering
  \includegraphics[width=0.9\linewidth]{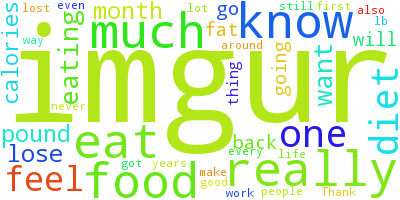}
  \caption{Top 10\%}
  \label{subfig:top10}
\end{subfigure}%
\begin{subfigure}{.5\linewidth}
  \centering
  \includegraphics[width=0.9\linewidth]{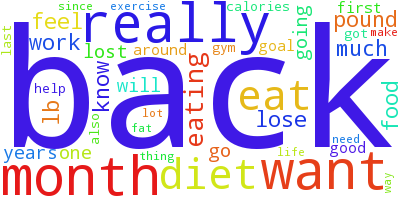}
  \caption{Bottom 10\%}
  \label{subfig:bottom10}
\end{subfigure}
\caption{Word clouds showing frequently used words in the most and least commented posts.}
\label{fig:wordClouds}
\vspace{-0.2cm}
\end{figure}

Examples of parts of highly-commented posts are ``Super \textbf{fat} and \textbf{really} gross me \url{http://i.imgur.com/ANON.jpg}'' 
 and ``\textbf{Feel} like sh*t so I \textbf{eat}, which makes me feel like sh*t, so I... and so on''. 
Bottom posts include passages such as ``I lost a lot of weight once but put it \textbf{back} on and much more. I'm finally ready to start again'' and ``I hit the \textbf{gym} and last year I hit 180. I've drifted \textbf{back} up to 190 since then''. 



Words present in the top posts are related to feelings, food and diet. Note that the most used word in this class is ``imgur''. This word is taken from an image sharing domain where people post before-after pictures showing their body transformation.\footnote{Shared pictures are hosted on the website \url{http://imgur.com}.} This suggests that\textit{ loseit} users feel that the community is a safe enough environment for them to share images of themselves in swim wear. Interestingly, the corresponding loss of anonymity does not seem to play a huge role for these sharing users and they do not seem to be fearing discrimination. Indeed, previous work has showed that this type of online groups can be particularly appealing to people with disabilities or social stigma, as they report mistreatment and discrimination because of their condition\cite{puhl:2001}. 
At the bottom end, there appear to be posts from users expressing their desire to get back on track and lose weight.




\section{Conclusion}
We presented a study on the initial posts of 37,278 users to the loseit weight loss subreddit. Applying Mahalanobis Distance Matching, we provide evidence that receiving any form of social feedback on the initial post, such as replying comments or upvotes, is linked to a roughly 18\% increase in the probability to return to the community for future activity. This link is not explained by variations in vocabulary of the original post. The boost in return probability is highest for feedback received on the first post, rather than later posts. The relative, additional gains in probability level off after receiving about 4-5 bits of feedback. Looking at the type of posts that are most likely to receive social feedback, we observe that posting images of oneself seems to be a good way to ensure a response from the community. Given the scarcity of studies on the effects of social feedback on continued engagement in health communities, we believe that these results contribute to a better understanding of the social dynamics underlying weight loss.

Looking back at our analysis, we were surprised to see a positive effect of receiving negative comments. Though this might be partly explained by an inadequacy of using VADER for this automatic ``supportive'' vs.\ ``discouraging'' labeling, more likely the loseit Reddit community is ``simply too nice''. In our study we did not observe any outright bitter or hateful comments. An example negative comment is 
``You f***ing skinny son of a b*tch... good on you mate.'', which despite the chosen terms is actually supportive. 
We are currently investigating the feasibility of looking at Twitter users who state their weight loss intention and to then look at the effect that receiving encouragement has on their inclination to continue posting updates. Given the prominence of abusive behavior such as ``fat shaming'' we would expect to see at least some negativity.

Lastly, we intend to go beyond studying the continued engagement in the weight loss community and, instead, focus on the effects on achieving actual weight loss. As a large number of community members regularly report their weight following conventions such as \emph{cw} for ``current weight'', we want to see if there is also a link between social feedback and weight loss.

%
\bibliographystyle{abbrv}
\bibliography{refs}  

\begin{thebibliography}{10}

\bibitem{Althoff:2014}
T.~Althoff, C.~Danescu{-}Niculescu{-}Mizil, and D.~Jurafsky.
\newblock How to ask for a favor: A case study on the success of altruistic
  requests.
\newblock {\em ICWSM}, 2014.

\bibitem{Ballantine:2011}
P.~W. Ballantine and R.~J. Stephenson.
\newblock {Help me, I'm fat! Social support in online weight loss networks}.
\newblock {\em Journal of Consumer Behaviour}, 2011.

\bibitem{Blei:2003}
D.~M. Blei, A.~Y. Ng, and M.~I. Jordan.
\newblock Latent dirichlet allocation.
\newblock {\em J. Mach. Learn. Res.}, 3, 2003.

\bibitem{Burke:2009}
M.~Burke, C.~Marlow, and T.~Lento.
\newblock Feed me: Motivating newcomer contribution in social network sites.
\newblock In {\em CHI}, 2009.

\bibitem{Chiu:2015}
C.-M. Chiu, H.-Y. Huang, H.-L. Cheng, and P.-C. Sun.
\newblock {Understanding online community citizenship behaviors through social
  support and social identity}.
\newblock {\em Int. Journal of Information Management}, 2015.

\bibitem{Choudhury:2014}
M.~D. Choudhury and S.~De.
\newblock Mental health discourse on reddit: Self-disclosure, social support,
  and anonymity.
\newblock In {\em ICWSM}, 2014.

\bibitem{eschler:2015}
J.~Eschler, Z.~Dehlawi, and W.~Pratt.
\newblock Self-characterized illness phase and information needs of
  participants in an online cancer forum.
\newblock In {\em ICWSM}, 2015.

\bibitem{Glasgow:2007}
R.~Glasgow, C.~Nelson, K.~Kearney, and \textit{et al}.
\newblock {{R}each, engagement, and retention in an {I}nternet-based weight
  loss program in a multi-site randomized controlled trial}.
\newblock {\em J. Med. Internet Res.}, 2007.

\bibitem{herzer:2011}
M.~Herzer, M.~Zeller, J.~Rausch, and A.~Modi.
\newblock Perceived social support and its association with obesity-specific
  health-related quality of life.
\newblock {\em Journal of developmental and behavioral pediatrics}, 2011.

\bibitem{Hutto:2014}
C.~J. Hutto and E.~Gilbert.
\newblock {VADER:} {A} parsimonious rule-based model for sentiment analysis of
  social media text.
\newblock In {\em ICWSM}, 2014.

\bibitem{king:2015}
G.~King and R.~Nielsen.
\newblock Why propensity scores should not be used for matching.
\newblock {\em http://j.mp/1sexgVw}, 2015.

\bibitem{Leroux:2013}
J.~S. Leroux, S.~Moore, and L.~Dube.
\newblock {{B}eyond the "{I}" in the obesity epidemic: a review of social
  relational and network interventions on obesity}.
\newblock {\em J Obes}, 2013.

\bibitem{patrick:2011}
K.~Patrick, K.~Calfas, G.~Norman, and \textit{et al}.
\newblock {{O}utcomes of a 12-month web-based intervention for overweight and
  obese men}.
\newblock {\em Ann Behav Med}, 2011.

\bibitem{puhl:2001}
R.~M. Puhl and K.~D. Brownell.
\newblock Bias, discrimination, and obesity.
\newblock {\em Obes Res}, 2001.

\bibitem{rosenbaum-1983}
P.~Rosenbaum and D.~Rubin.
\newblock The central role of the propensity score in observational studies for
  causal effects.
\newblock {\em Biometrika}, 70, 1983.

\bibitem{rubin:2006}
D.~B. Rubin and E.~A. Stuart.
\newblock Affinely invariant matching methods with discriminant mixtures of
  proportional ellipsoidally symmetric distributions.
\newblock {\em The Annals of Statistics}, pages 1814--1826, 2006.

\bibitem{tamersoy:2015}
A.~Tamersoy, M.~De~Choudhury, and D.~H. Chau.
\newblock Characterizing smoking and drinking abstinence from social media.
\newblock In {\em Hypertext}, 2015.

\bibitem{thiese:2015}
M.~Thiese, G.~Moffitt, R.~Hanowski, and \textit{et al}.
\newblock {{C}ommercial {D}river {M}edical {E}xaminations: {P}revalence of
  {O}besity, {C}omorbidities, and {C}ertification {O}utcomes}.
\newblock {\em J. Occup. Environ. Med.}, 2015.

\bibitem{umberson:2010}
D.~Umberson and J.~K. Montez.
\newblock Social relationships and health a flashpoint for health policy.
\newblock {\em Journal of health and social behavior}, 2010.

\bibitem{wang:2014}
M.~Wang, L.~Pbert, and S.~Lemon.
\newblock Influence of family, friend and coworker social support and social
  undermining on weight gain prevention among adults.
\newblock {\em Obesity}, 2014.

\end{thebibliography}
%
%

\end{document}